\documentclass[conference]{IEEEtran}
\IEEEoverridecommandlockouts

\def\finaldoi{}
\usepackage{eso-pic}
\newcommand{\acceptednote}{Accepted for publication at the International
  Conference on Artificial Intelligence Centric Systems, Services, and
  Applications (AISSA 2026), San Antonio, TX, USA, 8--10 December 2026.
  This is the authors' accepted manuscript.}
\newcommand{\ieeenotice}{\copyright~2026 IEEE. Personal use of this material
  is permitted. Permission from IEEE must be obtained for all other uses, in
  any current or future media, including reprinting/republishing this
  material for advertising or promotional purposes, creating new collective
  works, for resale or redistribution to servers or lists, or reuse of any
  copyrighted component of this work in other works.%
  \ifx\finaldoi\empty\else\ The final published version is available at
  \url{https://doi.org/\finaldoi}.\fi}
\AddToShipoutPictureBG*{%
  \AtPageUpperLeft{\raisebox{-\height}{\raisebox{-0.22in}{\makebox[\paperwidth]{%
    \parbox{\textwidth}{\centering\footnotesize\itshape\acceptednote}}}}}%
  \AtPageLowerLeft{\raisebox{0.38in}{\makebox[\paperwidth]{%
    \parbox[b]{\textwidth}{\centering\scriptsize\ieeenotice}}}}%
}

\usepackage{cite}
\usepackage{amsmath,amssymb}
\usepackage{graphicx}
\usepackage{booktabs}
\usepackage{url}
\usepackage[hidelinks]{hyperref}
\hypersetup{pdftitle={Trained on What Screen Readers Cannot Read: Accessibility Debt in the GUI Corpora Behind Computer-Use Agents},
            pdfauthor={Obada Kraishan and Kulsawasd Jitkajornwanich}}
\usepackage{siunitx}
\usepackage{tikz}
\usetikzlibrary{shapes.geometric,arrows.meta,positioning,calc}
\newcommand{\pct}[1]{#1\%}
\newcommand{\code}[1]{\texttt{#1}}

\begin{document}

\title{Trained on What Screen Readers Cannot Read: \\ Accessibility Debt in the GUI Corpora Behind Computer-Use Agents}

\author{\IEEEauthorblockN{Obada Kraishan}
\IEEEauthorblockA{\textit{College of Media and Communication}\\
\textit{Texas Tech University}\\
Lubbock, TX 79409, USA\\
omareikr@ttu.edu}
\and
\IEEEauthorblockN{Kulsawasd Jitkajornwanich}
\IEEEauthorblockA{\textit{College of Media and Communication}\\
\textit{Texas Tech University}\\
Lubbock, TX 79409, USA\\
kulsawasd.jitkajornwanich@ttu.edu}}

\maketitle

\begin{abstract}
Computer-use agents learn to operate mobile interfaces from corpora of
captured screens and accessibility trees. The same trees are what a screen
reader consumes, so an agent trained on them inherits whatever labelling gaps
the underlying applications carry. This paper measures one of those gaps,
content-description coverage, in two Android corpora: Rico, collected in 2017,
and AndroidControl, collected in 2024. Across \num{17373614} interface
elements, \pct{17.74} of the Rico elements that need a content description
under our coding rule carry one, against \pct{57.27} in AndroidControl. The
difference of \num{39.5} percentage points holds under five alternative coding
rules and when whole apps are resampled. It describes two datasets that also
differ in app selection, capture method and platform version; it does not
measure change over time. Input controls are the least labelled role in both
corpora (\pct{2.20} and \pct{27.21}, or \pct{30.99} in AndroidControl if hint
text counts as a name). Descriptions in the older corpus are also five times
more likely to be placeholders (\pct{10.25} against \pct{1.95}), so the
difference is not an artefact of developers typing filler. Resolving
\num{26205} task actions to the elements they touched shows that, among
tappable graphics with no text, demonstrations operate elements described
\num{9.0} points more often than the others on the screens they visit, about
\num{3.5} points of it from the choice of element within a screen. That is a
selection effect inside the benchmark. A two-coder reliability check
($\kappa = 0.91$ and $0.96$) and the full pipeline accompany the paper.
\end{abstract}

\begin{IEEEkeywords}
computer-use agents, GUI datasets, mobile accessibility, content descriptions,
benchmark validity, assistive technology
\end{IEEEkeywords}

\section{Introduction}

An agent that operates a phone on someone's behalf has to know what is on the
screen. The usual way to tell it is the platform accessibility tree: a
structured record of every element, its class, its bounds, whether it responds
to a tap, and its \code{content-desc}, the short name a developer supplies for
elements that show no text of their own. The large mobile GUI corpora used to
train and evaluate these agents are built from that tree. Rico~\cite{deka2017rico}
and AndroidControl~\cite{li2024androidcontrol} both expose it, and agents
built on AndroidControl, AndroidWorld and their relatives consume it
directly~\cite{li2024androidcontrol,rawles2025androidworld}.

The accessibility tree was built for screen readers, not for agents, and
developers have populated it poorly for as long as anyone has measured it
\cite{ross2020epidemiology,alshayban2020accessibility,fok2022longitudinal}.
That dependency has drawn little attention in the agent literature.
A missing content description makes an application harder to use with
TalkBack. It also removes information from the observation an agent is trained
on, evaluated on, and deployed against. We refer to this shared gap as
accessibility debt.

The framing changes who the problem belongs to. Missing descriptions have been
treated as an accessibility defect and addressed through developer tooling and
guidelines. If the corpora that train computer-use agents are drawn from the
same field, the defect is also a data-quality problem, and the agent research
community has both a reason and the resources to address it.

This paper measures the debt and asks four questions.

\begin{itemize}
\item \textbf{RQ1.} What share of the interface elements that require a
  content description carry one?
\item \textbf{RQ2.} How large is the difference in that share between a 2017
  corpus and a 2024 corpus?
\item \textbf{RQ3.} Do the elements that task demonstrations touch carry
  descriptions at a different rate from the other elements on the same
  screens?
\item \textbf{RQ4.} Is the share of described elements on an episode's screens
  associated with the length of its demonstration?
\end{itemize}

The paper makes four contributions. It provides a measurement of
content-description coverage across \num{17373614} elements in two corpora,
using a coding scheme that two coders applied independently with
$\kappa \geq 0.91$. It shows that the difference between corpora holds under
five alternative specifications, including one that gives the older corpus the
benefit of richer metadata and one that discounts uninformative descriptions,
and it reports the second of those as a finding in its own right. It resolves
task actions to the elements they touched and shows that demonstrations
favour better-described targets, which has a direct consequence for how agent
benchmarks should report results. Finally, it releases the codebook, the
pipeline and the statistical reports so that the same figures can be computed
for any corpus that exposes an accessibility tree.

The rest of the paper is organised as follows. Section~\ref{sec:related}
places the work in the context of GUI corpora, accessibility measurement and
benchmark validity. Section~\ref{sec:data} describes the two corpora and
Section~\ref{sec:method} the coding scheme, its reliability check and the
statistical approach. Section~\ref{sec:results} answers the four questions and
reports the sensitivity analysis. Sections~\ref{sec:discussion} to
\ref{sec:conclusion} discuss what the results mean for three audiences, state
the limitations, and conclude.

\section{Related Work}
\label{sec:related}

This section covers three strands of prior work: the corpora from which
interface agents learn, the measurement of missing labels in mobile
applications, and the question of what agent benchmarks measure. The paper
sits at their intersection.

\subsection{GUI corpora and the agents built on them}
Rico~\cite{deka2017rico} paired more than 66,000 screens from over 9,300
Android applications with their view hierarchies, and it underpinned a
generation of data-driven interface work, including widget
captioning~\cite{li2020widget}, screen
summarisation~\cite{wang2021screen2words} and the RICO-Semantics
annotations~\cite{sunkara2022semantics}. Later corpora attach task
demonstrations. Android in the Wild~\cite{rawles2023aitw} and
AndroidControl~\cite{li2024androidcontrol} record a person completing a
natural-language instruction, with the accessibility tree captured at each
step; AndroidWorld~\cite{rawles2025androidworld} turns the same idea into a
live benchmark. These are the resources that train and evaluate mobile
computer-use agents, and their observation format is inherited from the
platform rather than designed for the task. Some recent agents move away from
the tree toward pixels, in part because the tree is
incomplete~\cite{gou2025uground}; the present paper quantifies one form of
that incompleteness.

\subsection{Measuring missing labels in mobile applications}
Large-scale auditing of Android accessibility is well established. Ross et
al.\ applied an epidemiological framing to app accessibility and reported
widespread missing labels on image-based
buttons~\cite{ross2018buttons,ross2020epidemiology}. Fok et
al.~\cite{fok2022longitudinal} followed the same failure across app versions
over time. Alshayban et al.~\cite{alshayban2020accessibility} and Chen et
al.~\cite{chen2023accessibleornot} audited over a thousand apps each and
characterised both the defects and developer attitudes toward them, and
Yan and Ramachandran~\cite{yan2019status} surveyed the state of mobile
accessibility more broadly. Testing frameworks such as
Latte~\cite{salehnamadi2021latte} evaluate apps through the assistive
services a user would actually run. On the repair side, missing labels have
been predicted from component images~\cite{chen2020unblind}, icon classes
have been recovered at scale~\cite{chen2022iconlabeling}, and whole screens
have been reconstructed from pixels~\cite{zhang2021screenrecognition}. This
paper differs in what it measures rather than how. It examines the corpora
used to build agents, and it treats the missing labels as an input to a
learning pipeline rather than only as a barrier to a person.

\subsection{What agent benchmarks measure}
As benchmarks such as WebArena~\cite{zhou2024webarena},
OSWorld~\cite{xie2024osworld} and AndroidWorld~\cite{rawles2025androidworld}
become the field's yardstick, the question of what they actually measure
grows more consequential. RQ3 adds a specific validity concern: if the
elements a benchmark asks an agent to operate are better described than the
elements around them, then reported success is measured on targets that are
easier to identify than the rest of the screen.

\section{Data}
\label{sec:data}

This section describes the two corpora and the population of elements the
analysis draws from. The corpora were chosen because both expose the
accessibility tree and because they were collected seven years apart.

\textbf{Rico}~\cite{deka2017rico} contributes \num{66261} screens from a 2017
crawl of more than 9,300 free Android applications across 27 Play Store
categories, with view hierarchies and app-level metadata including category
and average rating. Rico is distributed under a research-use agreement that
forbids redistribution, so we release parsing code and checksums rather than
data.

\textbf{AndroidControl}~\cite{li2024androidcontrol} contributes \num{7623}
task episodes from ten of its twenty published shards (00000 to 00009),
covering \num{49822} screen observations, of which \num{26228} received a click or
long press (hereafter a click) that could be resolved to an element.
The data name an episode's app
only when the demonstration opens it, so we identify apps from the package
prefix of resource identifiers. This yields 700 application packages; 251 of
the episodes with elements needing a label carry no usable identifier.
Each step records an accessibility forest and an action. Actions carry no
element reference, only coordinates, so we resolve every click to an element
by point-in-bounds hit testing over the visible window stack, taking the
deepest node that contains the point and breaking ties on drawing order and
then on smallest area. When that node is not clickable, the click is assigned
to its nearest clickable ancestor; this applied to \pct{43.71} of resolved
clicks. A further \pct{18.42} stayed on a node that is not clickable, because
none of its ancestors was.
The rule resolved \pct{99.92} of clicks. Ten shards exceed every
sample-size threshold in the analysis plan by more than an order of
magnitude, so the remaining shards were not parsed. The release does not
group shards by app, split or date: the ten parsed shards hold between
\pct{48.95} and \pct{50.82} of every tenth of the released episode-identifier
range and \pct{50.75} of the released test episodes, and label coverage
(Section~\ref{sec:status}) ranges from \pct{55.84} to \pct{59.23} across
them.

Table~\ref{tab:corpora} summarises the analysis population that results once
the inclusion rule of Section~\ref{sec:unit} is applied.

\begin{table}[t]
\caption{Analysis population and label coverage by corpus.}
\label{tab:corpora}
\centering
\begin{tabular}{lrr}
\toprule
& \textbf{Rico 2017} & \textbf{AndroidControl 2024}\\
\midrule
Screens or step observations & \num{66261} & \num{49822}\\
Episodes & --- & \num{7623}\\
Elements parsed & \num{6371389} & \num{11002225}\\
In analysis population & \num{2862639} & \num{6831063}\\
Needs a label & \num{502585} & \num{1462544}\\
Labelled & \num{89146} & \num{837536}\\
\midrule
Label coverage & \pct{17.74} & \pct{57.27}\\
\quad 95\% CI, elements & $[17.63, 17.84]$ & $[57.19, 57.35]$\\
\quad 95\% CI, screen clusters & $[17.50, 17.97]$ & $[56.99, 57.53]$\\
\quad 95\% CI, app clusters & $[17.14, 18.37]$ & $[55.65, 58.86]$\\
Icon-like coverage & \pct{33.20} & \pct{62.75}\\
\bottomrule
\end{tabular}
\end{table}

\section{Method}
\label{sec:method}

This section defines the unit of analysis, the two coded variables, the strict
subpopulation used for the headline claims, the reliability procedure, and the
statistical approach. The scheme is implemented as code and stated in prose in
a codebook that ships with the pipeline.

\subsection{Unit of analysis and inclusion}
\label{sec:unit}
The unit is a single node in the view hierarchy (Rico) or accessibility tree
(AndroidControl). A node enters the analysis population when it is flagged
visible to the user and has positive area. Both conditions matter. In Rico,
\pct{48.7} of nodes have zero or negative area, and \num{1716} of those are
still flagged visible, so the visibility flag alone would admit elements that
are not on screen in any useful sense.
The \code{enabled} attribute is
not part of the primary rule, because a disabled control is still announced by
a screen reader and still needs a name. Section~\ref{sec:sensitivity} reports
the alternative.

\subsection{Element role}
Roles are assigned from the class name alone, matching on normalised suffixes
so that the rule survives AppCompat, Material, androidx and vendor subclasses.
Seven roles are used: \code{input}, \code{button}, \code{image}, \code{web},
\code{text}, \code{container} and \code{other}. Rules are evaluated in that
order and the first match wins, because several input classes end in
\code{-button} and one button class ends in \code{-imagebutton}.

Rico hierarchies carry an \code{ancestors} field with the full superclass
chain; AndroidControl trees do not. Classifying Rico with the richer signal
and AndroidControl with the poorer one would make any cross-corpus difference
partly an artefact of metadata rather than a fact about applications. The
primary classifier therefore uses the class name for both, and the superclass
variant is reported as a robustness check.

\subsection{Label status}
\label{sec:status}
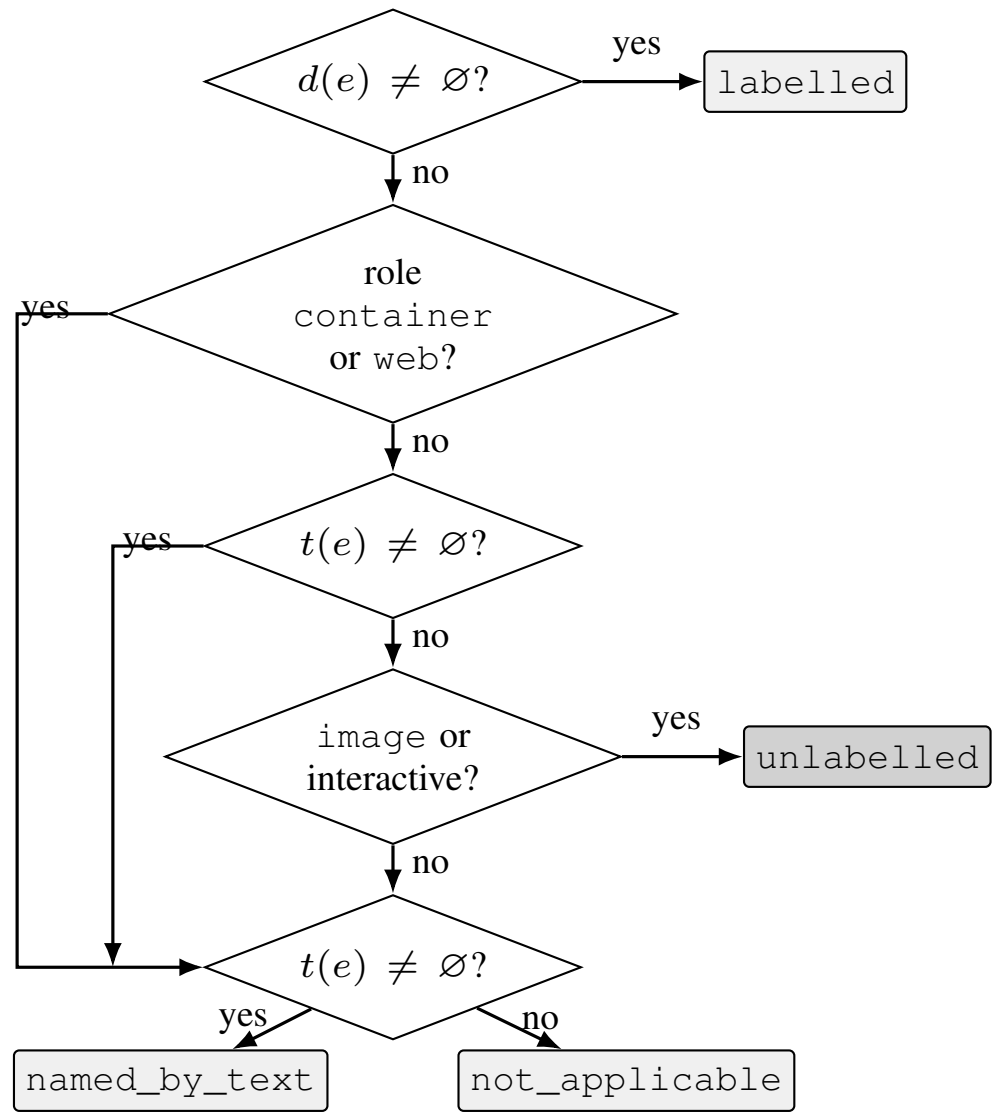
\begin{figure}[t]
\centering
\begin{tikzpicture}[
  font=\scriptsize,
  node distance=3.2mm and 4mm,
  q/.style={diamond,aspect=2.6,draw,inner sep=0.6pt,text width=17mm,align=center},
  r/.style={rectangle,draw,rounded corners=1pt,inner sep=2.2pt,minimum height=4mm},
  lab/.style={r,fill=black!6},
  deb/.style={r,fill=black!18},
  arr/.style={-{Latex[length=1.6mm]},semithick}
]
\node[q] (q1) {$d(e)\neq\varnothing$?};
\node[q,below=of q1] (q2) {role \code{container} or \code{web}?};
\node[q,below=of q2] (q3) {$t(e)\neq\varnothing$?};
\node[q,below=of q3] (q4) {\code{image} or interactive?};
\node[q,below=of q4] (q5) {$t(e)\neq\varnothing$?};
\node[lab,right=8mm of q1] (l) {\code{labelled}};
\node[deb,right=8mm of q4] (u) {\code{unlabelled}};
\node[lab,below left=3mm and -2mm of q5] (n) {\code{named\_by\_text}};
\node[lab,below right=3mm and -2mm of q5] (na) {\code{not\_applicable}};
\draw[arr] (q1) -- node[above,pos=.45]{yes} (l);
\draw[arr] (q1) -- node[right,pos=.4]{no} (q2);
\draw[arr] (q2) -- node[right,pos=.4]{no} (q3);
\draw[arr] (q3) -- node[right,pos=.4]{no} (q4);
\draw[arr] (q4) -- node[above,pos=.45]{yes} (u);
\draw[arr] (q4) -- node[right,pos=.4]{no} (q5);
\draw[arr] (q5) -- node[left,pos=.3]{yes} (n);
\draw[arr] (q5) -- node[right,pos=.3]{no} (na);
\draw[arr] (q2.west) -- node[left,pos=.2]{yes} ++(-6mm,0) |- (q5.west);
\draw[arr] (q3.west) -- node[left,pos=.2]{yes} ++(-4mm,0) -| ($(q5.west)+(-6mm,0)$);
\end{tikzpicture}
\caption{The label-status decision. Questions 2 and 3 route exempt elements
past the debt test to the final split. The shaded outcome, \code{unlabelled},
is the accessibility debt.}
\label{fig:tree}
\end{figure}

Each included element receives exactly one of four values, assigned by the
decision procedure in Fig.~\ref{fig:tree}. Let $d(e)$ be the element's
content description after whitespace and the placeholder strings
\code{none}, \code{null} and \code{nan} are removed, and let $t(e)$ be its
visible text treated the same way. An element \emph{needs} a description when
\begin{equation}
\label{eq:needs}
\begin{split}
\mathrm{needs}(e) = {}& \neg\mathrm{exempt}(e) \wedge t(e) = \varnothing\\
  & {} \wedge \big(\mathrm{image}(e) \vee \mathrm{interactive}(e)\big),
\end{split}
\end{equation}
where $\mathrm{exempt}(e)$ holds for roles \code{container} and \code{web},
and $\mathrm{interactive}(e)$ holds when the element is clickable or
long-clickable or has role \code{input} or \code{button}. Containers are
structural and are traversed rather than announced; a web view hides its
contents from the tree; an element that shows text is announced by that text;
and an image always needs a description because a screen reader has nothing
else to say.

An element is \code{labelled} when $d(e) \neq \varnothing$, whether or not it
also shows text. It is \code{unlabelled} when $d(e) = \varnothing$ and
$\mathrm{needs}(e)$ holds; this is the debt. Otherwise it is
\code{named\_by\_text} if it shows text and \code{not\_applicable} if it shows
nothing. Coverage for any population $P$ is then
\begin{equation}
\label{eq:coverage}
C(P) = \frac{\left|\{e \in P : \mathrm{needs}(e) \wedge d(e) \neq \varnothing\}\right|}
            {\left|\{e \in P : \mathrm{needs}(e)\}\right|}.
\end{equation}
The denominator is the needs-a-label population, never the count of elements
coded \code{labelled} or \code{unlabelled}. The latter would report exempt
roles at \pct{100} by construction, since no container is ever coded
\code{unlabelled}.

The construct measured throughout is therefore content-description coverage
under the stated rule. It is not the same as whether a screen reader can name
an element, because TalkBack can also take an accessible name from other
fields, most often the hint text of an input field;
Section~\ref{sec:rq1} reports coverage with hint text counted.

\subsection{The icon-like subpopulation}
We define an element as \emph{icon-like} when it is a tappable graphic with
no text of its own: role \code{image} or \code{button}, $t(e) = \varnothing$,
and $\mathrm{interactive}(e)$. This is the strictest unit in the paper and the
population where a missing description is a defect rather than a judgement
call. Every headline claim is restated on it.

\subsection{Reliability}
\label{sec:reliability}
Two authors independently hand-coded a stratified sample of 300 elements (60
screens per corpus, up to 10 elements per screen, stratified by role and by
automated label status so that rare cells are represented), each blind to the
other and to the automated assignments. Agreement was measured with Cohen's
$\kappa$~\cite{cohen1960kappa},
\begin{equation}
\label{eq:kappa}
\kappa = \frac{p_o - p_e}{1 - p_e},
\end{equation}
where $p_o$ is observed agreement and $p_e$ is agreement expected by chance,
with percentile intervals from \num{2000} bootstrap resamples of the coded
items and the Landis and Koch conventions for
interpretation~\cite{landis1977kappa}.

The two coders reached $\kappa = 0.96$, \pct{95} CI $[0.94, 0.98]$ and
\pct{97.0} raw agreement for element role, and $\kappa = 0.91$, \pct{95} CI
$[0.87, 0.95]$ and \pct{93.3} raw agreement for label status. The 28 items on
which they differed were not adjudicated: the human codes serve only to
check the scheme and enter no analysis, and every result below rests on the
automated assignment. Agreement was lowest at the
boundary between \code{not\_applicable} and \code{unlabelled}, which accounted
for 11 of the 20 label-status disagreements. That boundary is where the
decorative-versus-meaningful judgement sits, and Section~\ref{sec:limitations}
returns to it.

Because the automated coder is deterministic and cannot disagree with itself,
its agreement with each human coder is reported as a validity check rather
than as reliability: $\kappa = 0.98$ and $0.99$ for role, and $0.96$ and
$0.95$ for label status. Eleven automated validation checks on vocabularies,
screen geometry and internal contradictions passed without exception.

\subsection{Statistical approach}
Proportions carry Wilson score intervals~\cite{wilson1927}, which stay
well-behaved at the low rates seen in the older corpus. Differences between
proportions carry Newcombe hybrid score intervals~\cite{newcombe1998} and are
accompanied by Cohen's $h$~\cite{cohen1992power},
\begin{equation}
\label{eq:h}
h = \left|\,2\arcsin\sqrt{p_1} - 2\arcsin\sqrt{p_2}\,\right|,
\end{equation}
reported as a magnitude with direction given in the text and read against the
conventions of 0.20, 0.50 and 0.80 for small, medium and large. Contingency
tables carry Cram\'er's $V$ with bootstrap intervals, and ordinal comparisons
carry Cliff's $\delta$~\cite{cliff1993}. RQ3 uses permutation tests with
\num{10000} label shuffles and the $(r+1)/(m+1)$ convention, alongside
parametric companions. Because elements are nested in screens and screens in
apps, the headline proportions, the cross-corpus difference and the RQ3
differences also carry percentile intervals from \num{5000} cluster-bootstrap
resamples of whole screens and, for coverage and the difference, of whole
apps. RQ3 adds a paired
within-screen difference and a permutation test that shuffles the touched
flag only within screens. The null of that test is centred on the difference
that screen composition alone would produce, so it tests the part of each
difference that arises within screens.
Where a family of tests is reported, $p$-values are
adjusted by the Benjamini--Hochberg procedure~\cite{benjamini1995fdr}. RQ4
fits a negative binomial model of demonstration length,
\begin{equation}
\label{eq:nb}
\log \mathbb{E}[\mathrm{steps}_i] = \beta_0 + \beta_1\,\mathrm{density}_i +
  \beta_2 \log n_i ,
\end{equation}
where $\mathrm{density}_i$ is the share of labellable elements on episode
$i$'s screens that carry a description and $n_i$ is the total number of
included elements across those screens, so that $\exp(\beta_1)$ is an
incidence rate ratio for label density at a fixed element count. Because
$n_i$ is summed over the screens an episode visits, it grows with the number
of steps. Model~2 adds AndroidControl's published split (train, validation,
test) and clusters standard errors by app. Model~3 keeps the split and the
clustering but takes both density and element count from the episode's first
in-app screen, defined as observation 1 after an \code{open\_app} action and
observation 0 otherwise, for episodes where that screen has at least five
elements needing a label. Exposure and control are then fixed before the
demonstration unfolds. All analyses are seeded, and every
number in this paper can be traced to a generated report.

\section{Results}
\label{sec:results}

\begin{figure*}[t]
\centering
\includegraphics[width=\textwidth]{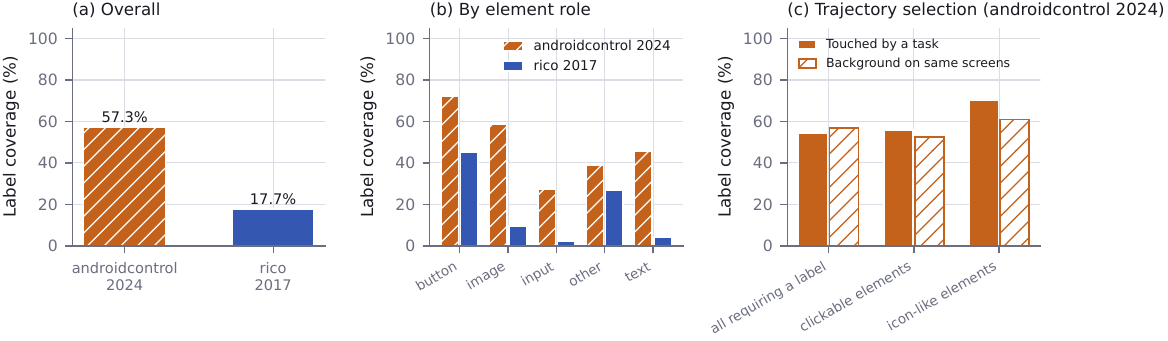}
\caption{Label coverage across the two corpora. (a) Overall coverage among
elements requiring a description. (b) Coverage by element role; exempt roles
are omitted. (c) Within AndroidControl, elements touched by a task
demonstration against the background population on the same screens. Hatching
duplicates the colour encoding so the figure reads in greyscale.}
\label{fig:main}
\end{figure*}

This section answers the four research questions in turn and then reports how
far the headline figures move under alternative coding rules.
Fig.~\ref{fig:main} gives the overall picture.

\subsection{RQ1: coverage of the label surface}
\label{sec:rq1}

Among elements that require a description, coverage is \pct{17.74}
(\pct{95} CI $[17.63, 17.84]$) in Rico and \pct{57.27} ($[57.19, 57.35]$) in
AndroidControl. Resampling whole apps widens these intervals to
$[17.14, 18.37]$ and $[55.65, 58.86]$ (Table~\ref{tab:corpora}), and the two
estimates stay far apart. Both are lower bounds. The tree cannot tell a
decorative image from a meaningful one, so some elements counted as needing
a description need none, and that boundary is where the two coders disagreed
most (Section~\ref{sec:reliability}). On the icon-like subpopulation the figures are \pct{33.20}
($[32.94, 33.45]$) and \pct{62.75} ($[62.54, 62.97]$). Icon-like elements are
better described than other elements needing a label in both corpora, and the
gap is medium in Rico (risk difference \num{20.7} points $[20.4, 21.0]$,
$h = 0.51$) but small in AndroidControl (\num{6.3} points $[6.1, 6.6]$,
$h = 0.13$).

Label status depends on element role. The association is medium in Rico,
$\chi^2(4) = \num{63501.8}$, $p < .001$, $V = 0.36$ $[0.35, 0.36]$, and
small in AndroidControl, $\chi^2(4) = \num{36466.6}$, $p < .001$, $V = 0.16$
$[0.16, 0.16]$. Table~\ref{tab:roles} gives the breakdown together with the
cross-corpus difference for each role.

\begin{table}[t]
\caption{Label coverage by element role and the cross-corpus difference.}
\label{tab:roles}
\centering
\begin{tabular}{lS[table-format=2.2]S[table-format=2.2]S[table-format=+2.2]S[table-format=1.2]}
\toprule
\textbf{Role} & {\textbf{Rico}} & {\textbf{AndroidControl}} & {\textbf{Diff}} & {$\boldsymbol{h}$}\\
 & {(\%)} & {(\%)} & {(pp)} & \\
\midrule
\code{button} & 45.28 & 72.26 & +26.98 & 0.56\\
\code{image}  &  9.62 & 58.42 & +48.81 & 1.11\\
\code{input}  &  2.20 & 27.21 & +25.01 & 0.80\\
\code{text}   &  4.09 & 45.52 & +41.43 & 1.07\\
\code{other}  & 26.63 & 38.91 & +12.28 & 0.26\\
\bottomrule
\end{tabular}
\end{table}

The \code{input} row deserves attention. Text fields, checkboxes, switches and
pickers are the least described role in both corpora, at \pct{2.20} in Rico
and \pct{27.21} in AndroidControl. The elements a person must enter data into
are the ones least likely to have a name, in the category where the cost of
guessing wrong is highest. For an input, the main alternative source of a
spoken name is its hint text. Rico's hierarchies do not record hints;
AndroidControl's do, and counting a non-empty hint as a name raises input
coverage there from \pct{27.21} to \pct{30.99}. Inputs remain the least
described role, and the same change moves image and button coverage by less
than a tenth of a point.

App-level metadata, available for Rico only, explains little. Across 27 Play
Store categories, all differences from the pooled rate are below the small
threshold for $h$ except House and Home (\pct{28.14}, \num{10.5} points above
the rest, $h = 0.25$) and Weather (\pct{7.78}, \num{10.6} points below,
$h = 0.32$). Average rating is associated with coverage at
$\chi^2(3) = 88.09$, $p < .001$, but the effect is negligible,
$V = 0.013$ $[0.011, 0.016]$; coverage runs from \pct{19.38} in apps rated
below 3.5 to \pct{17.61} in apps rated above 4.5. At these sample sizes a
small $p$-value carries almost no information on its own. The practical
reading is that highly rated apps are not more accessible than poorly rated
ones, which suggests users do not perceive this class of defect, or do not
penalise it.

\subsection{RQ2: the difference between corpora}

Coverage is \num{39.53} percentage points higher in the 2024 corpus,
$z = 484.29$, $p < .001$, risk difference $0.395$ $[0.394, 0.397]$,
$h = 0.85$, a large effect on $N = \num{1965129}$ elements. Resampling whole
screens gives an interval of $[39.17, 39.90]$ points and resampling whole apps
$[37.81, 41.21]$. The difference holds in every role (Table~\ref{tab:roles})
and on the icon-like subpopulation, where it is \num{29.6} points
$[29.2, 29.9]$, $h = 0.60$. It is largest for \code{image} elements, at
\num{48.8} points with $h = 1.11$, the role where a missing description leaves
a screen reader with nothing at all to announce.

We read the difference as one between datasets, not as change over time. The
two corpora differ in app selection, capture method and platform version as
well as in collection year. The AndroidControl release covers 833 applications
chosen to support task demonstrations~\cite{li2024androidcontrol}, a narrower
and probably better-maintained population than Rico's broad crawl, and that
alone would push its figure upward. What the comparison does
establish is that the corpus the field currently trains on carries much less
debt than the corpus it trained on for most of the past decade, while still
leaving \pct{42.7} of the elements that need a description without one.

\subsection{RQ3: which elements task demonstrations touch}

AndroidControl records the element a demonstrator touched only implicitly,
through coordinates. Resolving \num{26205} touched elements across
\num{26228} screens and \num{7342} episodes lets us compare them with the
\num{3556428} background elements on the same screens. Restricting the
comparison to the same screens matters: comparing touched elements with the
whole corpus would confound which elements tasks touch with which screens
tasks visit.

Table~\ref{tab:rq3} gives the three comparisons. On all elements requiring a
label, touched elements are described \emph{less} often than the background,
\pct{54.14} against \pct{56.88}, a difference of \num{-2.7} points
$[-3.9, -1.5]$. On clickable elements only, the direction reverses to
\num{+3.3} points $[2.1, 4.5]$. On icon-like elements the advantage is
\num{+9.0} points $[7.5, 10.6]$, $h = 0.19$, with the permutation and
parametric tests agreeing throughout. Resampling whole screens gives
$[7.5, 10.5]$ for the icon-like difference, and every row keeps its sign
(Table~\ref{tab:rq3}).

\begin{table}[t]
\caption{Label coverage of touched versus background elements on the same screens.}
\label{tab:rq3}
\centering
\setlength{\tabcolsep}{2.5pt}
\begin{tabular}{lS[table-format=2.2]S[table-format=2.2]S[table-format=+1.2]cS[table-format=+1.2]}
\toprule
\textbf{Population} & {\textbf{Touched}} & {\textbf{Backgr.}} & {\textbf{Diff}} & {\textbf{Screen-cluster}} & {\textbf{Within}}\\
 & {(\%)} & {(\%)} & {(pp)} & {\textbf{95\% CI}} & {(pp)}\\
\midrule
All needing a label & 54.14 & 56.88 & -2.74 & $[-3.95, -1.47]$ & -0.80\\
Clickable only      & 55.90 & 52.59 & +3.31 & $[2.15, 4.51]$   & +1.99\\
Icon-like only      & 69.96 & 60.97 & +8.99 & $[7.47, 10.51]$  & +3.52\\
\bottomrule
\end{tabular}
\end{table}

Shuffling the touched flag only within screens separates two sources of each
difference: where touches fall, and which element is touched on a screen. The
last column of Table~\ref{tab:rq3} gives the within-screen part; the rest
comes from screen composition. For icon-like elements, about \num{5.5} of the
\num{9.0} points arise because touches concentrate on screens whose icons are
better described, and about \num{3.5} because the touched element is better
described than the others on its own screen. The paired within-screen
difference is \num{+4.5} points $[3.4, 5.6]$, and the stratified test gives
$p < .001$. Both parts are selection effects inside the benchmark.

The first row must be read with its denominator in mind. Most touched elements
are clickable (Section~\ref{sec:data}), whereas the background for that row is dominated by
non-interactive images, described at whatever rate images are described. The
row compares two different populations, and the decomposition bears this
out: its within-screen part is \num{-0.8} points and the stratified test gives
$p = .14$, so the negative difference comes mostly from which screens tasks
visit. The second and third rows hold interactivity fixed, and both point the
same way.

Two checks bear on the hit-testing rule. Touched icon-like elements are
slightly larger than background ones (median \num{17424} against \num{16002}
square pixels, Cliff's $\delta = 0.176$ $[0.157, 0.194]$, a small effect).
Restricting to direct hits, which drops the \pct{43.71} of clicks assigned to
a clickable ancestor, leaves the icon-like difference at \num{+7.43} points
$[5.77, 9.06]$; ancestor-assigned icon-like touches are described more often
than direct ones (\pct{77.26} against \pct{68.40}). Direct hits also give
\num{+10.06} points for clickable elements and \num{+2.65} for all elements,
so the ancestor rule does not manufacture the effect.

Read on the strictest unit, then, the elements a benchmark asks an agent to
operate carry descriptions more often than the tappable graphics it does not
ask about, both across the screens tasks visit and within a single screen.
Demonstrations were produced by people completing goals, and people may reach
for the affordances they can identify. Whether this makes reported success
optimistic depends on an assumption these data cannot test: that the
elements a deployed agent must operate resemble the untouched background more
than the touched targets. The effect is modest in size; its direction is the
finding.

\subsection{RQ4: label density and demonstration length}

Described without adjustment, episodes below the median label density of
$0.506$ have longer demonstrations: mean steps are $5.71$ $[5.58, 5.83]$
below the median and $5.39$ $[5.25, 5.52]$ above it, a difference of $0.32$
steps $[0.13, 0.50]$, Cliff's $\delta = 0.06$.

Adjusting for element count with the model in \eqref{eq:nb} is not a clean
test, because $n_i$ is summed over the screens a demonstration visits and so
grows with its length: its correlation with log steps is $0.80$, and its own
rate ratio is $2.25$. Model~3 in Table~\ref{tab:rq4} answers the question
with exposure and control fixed before the demonstration unfolds, taking both
from the first screen inside the app for the \num{7407} episodes where that
screen has at least five elements needing a label. Under this specification
there is no association: the rate ratio for label density is $0.986$
$[0.895, 1.086]$, $p = .78$. Models~1 and~2 show an association only because
they condition on a quantity that grows with the outcome. For the same
reason we do not enter the number of screens visited, which equals steps
plus one in all \num{7615} episodes because every step records one
observation.

\begin{table}[t]
\caption{Incidence rate ratio for label density in three models of demonstration length.}
\label{tab:rq4}
\centering
\setlength{\tabcolsep}{4pt}
\begin{tabular}{clrc}
\toprule
& \textbf{Density and $n$ from} & \textbf{N} & \textbf{IRR [95\% CI]}\\
\midrule
1 & All screens                         & \num{7615} & $0.818$ $[0.783, 0.855]$\\
2 & All screens; split, clustered SE    & \num{7615} & $0.819$ $[0.740, 0.906]$\\
3 & First in-app screen; split, cl.\ SE & \num{7407} & $0.986$ $[0.895, 1.086]$\\
\bottomrule
\end{tabular}
\end{table}

The answer to RQ4 is therefore that these data give no evidence of an
association between label density and demonstration length once exposure and
control are fixed before the demonstration. Two further cautions apply. The
outcome is demonstration length, not agent success, since AndroidControl
contains human demonstrations and no agent rollouts. And length is a coarse
proxy for difficulty, since a long demonstration may reflect a complex goal
rather than a hard interface.

Label density is close to flat across the benchmark's four generalisation
splits (Table~\ref{tab:splits}). Split labels exist only for AndroidControl's
test episodes, and the four splits overlap, so the table's rows hold
\num{1432} memberships across 780 distinct test episodes, while Models~1
and~2 use all \num{7615} episodes. The association is detectable,
$\chi^2(3) = 38.08$, $p < .001$, but negligible in size, $V = 0.013$
$[0.009, 0.017]$. This is a useful null. Whatever makes the unseen-app split
harder for an agent, it is not that its screens are less well described.

\begin{table}[t]
\caption{Label density and mean demonstration length across AndroidControl's
generalisation splits.}
\label{tab:splits}
\centering
\begin{tabular}{lS[table-format=3]S[table-format=2.2]S[table-format=1.2]}
\toprule
\textbf{Split} & {\textbf{Episodes}} & {\textbf{Coverage (\%)}} & {\textbf{Mean steps}}\\
\midrule
In distribution   & 364 & 55.17 & 5.37\\
Unseen category   & 349 & 56.20 & 5.77\\
Unseen app        & 318 & 54.78 & 5.74\\
Unseen task       & 401 & 56.26 & 5.72\\
\bottomrule
\end{tabular}
\end{table}

\subsection{Sensitivity to coding decisions}
\label{sec:sensitivity}

Every figure above rests on choices a reader can reasonably question. Rather
than defend one specification, Table~\ref{tab:sens} recomputes the headline
difference under five.

\begin{table}[t]
\caption{Cross-corpus coverage difference under five specifications.}
\label{tab:sens}
\centering
\begin{tabular}{clS[table-format=2.2]S[table-format=2.2]S[table-format=+2.2]S[table-format=1.2]}
\toprule
& \textbf{Specification} & {\textbf{Rico}} & {\textbf{AC}} & {\textbf{Diff}} & {$\boldsymbol{h}$}\\
& & {(\%)} & {(\%)} & {(pp)} & \\
\midrule
A & Primary                       & 17.74 & 57.27 & +39.53 & 0.85\\
B & Exclude disabled elements     & 17.77 & 57.30 & +39.54 & 0.85\\
C & Icon-like only                & 33.20 & 62.75 & +29.56 & 0.60\\
D & Rico via superclass chain     & 19.36 & 57.27 & +37.91 & 0.81\\
E & Uninformative labels absent   & 15.92 & 56.15 & +40.23 & 0.87\\
\bottomrule
\end{tabular}
\end{table}

The difference never changes sign and never collapses. Its narrowest form is
variant C, the strictest unit, at \num{29.6} points.

Variant D addresses the most direct objection to RQ2, that Rico's figure is
low because our classifier saw less of its metadata. Reclassifying Rico with
its full superclass chain while leaving AndroidControl untouched changes the
role of \num{320059} Rico elements (\pct{11.18} of its included population)
and moves its coverage from \pct{17.74} to \pct{19.36}. The difference falls
by \num{1.6} points, from \num{39.53} to \num{37.91}. Unequal metadata
accounts for about \pct{4} of the gap.

Variant E addresses the opposite objection, that the 2024 figure is inflated
by placeholder strings. We count a present description as uninformative when
it names the widget rather than its function (``image'', ``button''), echoes
the element's resource identifier, or is a single character or a bare number.
Such strings make up \pct{1.95} of AndroidControl's descriptions
(\num{16310} of \num{837536}) but \pct{10.25} of Rico's (\num{9141} of
\num{89146}). The older corpus did not only describe fewer elements; the
descriptions it did have were five times more likely to be placeholders.
Counting them as absent widens the difference to \num{40.23} points. The
primary specification therefore understates the difference between the
corpora rather than inflating it.

\section{Discussion}
\label{sec:discussion}

This section draws out what the results mean for three groups: the people who
build GUI corpora, the people who design agent benchmarks, and the people who
build agents. It closes with the shared interest that connects them.

\subsection{For corpus builders}
Coverage is a property of a dataset and should be published with one. A
corpus card that reported label coverage overall, by role, and on the icon-like
subpopulation would let a reader see how much of the observation space is
described before training on it. The codebook and pipeline released with this
paper compute those figures from any accessibility tree.

\subsection{For benchmark designers}
RQ3 shows that trajectory-derived benchmarks inherit a selection effect from
the people who produced the trajectories. The remedy is not to discard the
data but to report against it: publish agent success stratified by whether
the target element carries a description. An agent that succeeds only on
described targets and one that succeeds on both are different systems, and an
aggregate success rate cannot tell them apart. Sampling task targets to
balance described and undescribed elements would go further.

\subsection{For agent builders}
An agent that consumes the accessibility tree consumes an artefact of
developer diligence. About \pct{43} of the elements needing a description in
the newer corpus have none, and for input controls the figure is \pct{73}, or
\pct{69} if hint text counts as a name.
Systems that rely on the tree alone therefore inherit a blind spot of that
size, which is a concrete argument for treating pixel-based recovery of
interface semantics~\cite{zhang2021screenrecognition,chen2020unblind,gou2025uground}
as a complement to the tree rather than as an alternative to it.

\subsection{A shared interest}
Accessibility researchers have argued for populated content descriptions for a
decade on the grounds that screen reader users need them. The agent research
community now needs the same field populated for a different reason.
Improving it serves both, and the second group brings resources and commercial
incentives the first has rarely had.

\section{Limitations}
\label{sec:limitations}

This section states the boundaries of the evidence. The view hierarchy cannot
distinguish a decorative image from a meaningful one, so elements coded
\code{unlabelled} include some that need no description. Every coverage figure
is therefore a lower bound and every debt figure an upper bound. That is the
conservative direction for the paper's claims, but it is also the largest
single source of measurement error, and it is where the two human coders
disagreed most (Section~\ref{sec:reliability}). Resolving it requires a
judgement the tree cannot supply, namely whether a person could name the
element's function. Human icon annotations such as those in
RICO-Semantics~\cite{sunkara2022semantics} make that judgement available for
part of Rico, and applying them is the natural next step.

Coverage counts content descriptions only. A screen reader can also announce
hint text, an associated label or the text of child elements, so an element
without a description is not always silent. The hint-text variant measures
the largest of these for inputs in AndroidControl; Rico records no hints.

Elements are nested within screens and screens within apps, so they are not
independent observations. The headline coverage figures, the cross-corpus
difference and RQ3 therefore also carry screen- and app-level cluster-bootstrap
intervals, which are wider but change no conclusion. The contingency tests in
RQ1 still treat elements as independent and should be read as descriptive.

Touched elements in RQ3 are inferred from coordinates. A systematic bias in
the hit-testing rule, such as a preference for larger targets, would
propagate into the result. The \pct{99.92} resolution rate, the
clickable-only comparison, the size comparison and the restriction to direct
hits guard against it, and none of them reverses the result. The
uninformative-description rule in variant E is a keyword list, not a semantic
judgement, and it will miss a description that is fluent but wrong; it is a
lower bound on low-quality labelling. Finally, the two corpora differ in more
ways than their capture dates, so RQ2 compares datasets and does not measure
change over time.

\section{Conclusion}
\label{sec:conclusion}

Computer-use agents see interfaces through a channel built for screen readers,
and they inherit its gaps. Measuring \num{17373614} elements across a 2017 and
a 2024 GUI corpus, this paper finds label coverage of \pct{17.74} and
\pct{57.27}, a difference that holds under every alternative rule we tried and
that is understated rather than inflated once placeholder descriptions are
discounted. The difference describes two datasets that also differ in app
selection, capture method and platform; it does not measure change over time.
Input controls are the worst-served role in both corpora. And the elements
task demonstrations touch are better described than the other tappable
graphics on the screens they visit, by nine points, of which about
\num{3.5} come from the choice of element within a screen. That is a
selection effect inside the benchmark; whether it makes reported performance
optimistic depends on what deployed agents are asked to operate.
Accessibility debt is not only a
barrier to disabled users. It is a measurable defect in the training and
evaluation data of the agents now being built to use interfaces on everyone's
behalf, and it can be measured with the tools released alongside this paper.

\section*{Data and Code Availability}
The codebook, the analysis pipeline, the generated statistical reports and
the file that traces every number in this paper to its report are available
at \url{https://github.com/obadaKraishan/accessibility-debt-agents}. Rico is
distributed under a research-use agreement and is not redistributed; the
repository provides parsing code and checksums instead. AndroidControl is
available from its authors under the Apache 2.0 licence.

\bibliographystyle{IEEEtran}
\bibliography{refs}

\end{document}